\documentclass[aip,amsmath,amssymb,reprint]{revtex4-1}
\usepackage[english]{babel}
\usepackage{graphicx}
\usepackage{dcolumn}
\usepackage{bm}

 \usepackage[dvipdfmx]{color}

\newcommand{\PhiL}{\Phi_{L}(k,t)}
\newcommand{\PhiS}{\Phi_{S}(k,t)}
\newcommand{\PhisL}{\Phi_{s, L}(k,t)}
\newcommand{\PhisS}{\Phi_{s, S}(k,t)}
\newcommand{\PhicL}{\Phi^{c}_{L}(k,t)}
\newcommand{\PhicS}{\Phi^{c}_{S}(k,t)}
\newcommand{\PhicsL}{\Phi^{c}_{s, L}(k,t)}
\newcommand{\PhicsS}{\Phi^{c}_{s, S}(k,t)}
\newcommand{\tPhisL}{\widetilde{\Phi}_{s, L}(k,t)}
\newcommand{\tPhisS}{\widetilde{\Phi}_{s, S}(k,t)}
\newcommand{\lgle}{\left\langle}
\newcommand{\rgle}{\right\rangle}
\newcommand{\vk}{\tilde{v}(k)}
\newcommand{\pdif}[2]{\frac{\partial #1}{\partial #2 } }
\newcommand{\Hk}[1]{H_{k}^{(#1)}}

\begin{document}

\title{Slow dynamics coupled with cluster formation in ultrasoft-potential glasses}

\author{Ryoji Miyazaki}
\altaffiliation[Present address: ]{Graduate School of Information Science, Tohoku University, Sendai 980-8579, Japan}
\author{Takeshi Kawasaki}
\author{Kunimasa Miyazaki}
\email[Corresponding author: ]{miyazaki@r.phys.nagoya-u.ac.jp}
\affiliation{Department of Physics, Nagoya University, Nagoya, Japan}

\date{\today}

\begin{abstract}
We numerically investigate slow dynamics of a binary mixture of ultrasoft particles 
interacting with the generalized Hertzian potential. 
If the softness parameter, $\alpha$, is small, the particles at high densities start penetrating each
other, form clusters, and eventually undergo the glass transition. 
We find multiple cluster-glass phases characterized by different number of particles per cluster, whose
boundary lines are sharply separated by the cluster size. 
Anomalous logarithmic slow relaxation of the density correlation functions 
is observed in the vicinity of these glass-glass phase boundaries, which
hints the existence of the higher-order dynamical singularities predicted by the mode-coupling theory. 
Deeply in the cluster glass phases, it is found that the dynamics of a single particle is decoupled
from that of the collective fluctuations. 
\end{abstract}


\maketitle

\section{Introduction}

The ultrasoft potentials are pairwise isotropic and repulsive interactions whose value remains 
finite even if the two particles fully overlap each other~\cite{C.Likos2006, G.Malescio2007}. 
The Gaussian, harmonic, and Hertzian potentials are typical examples.
The ultrasoft potential systems are known to exhibit very rich and complex thermodynamic and dynamic
behaviors at high densities, 
which are never observed in systems with sharp short-ranged repulsions such as the hard-sphere and Lennard-Jones potentials.
As originally proposed by Likos {\it et al.} based on a mean-field analysis, 
the ultrasoft potential systems are categorized crudely into two types, {\it i.e.}, the $Q^{+}$ and
$Q^{\pm}$ classes, depending on the shape of the interparticle potential $v(r)$, or more precisely,
the shape of its Fourier transformation $\vk$~\cite{C.Likos2001}. 
The $Q^{+}$ class is a system with a positive-definite $\vk$, where $\vk >0$ for all wavevectors $k$. 
The Gaussian-core model (GCM) and Hertzian potential system belong to this class~\cite{F.Stillinger1976}.  
The most salient feature of the systems in this class is the reentrant melting. 
The system which crystallized at moderate densities at a fixed temperature melts again as the
density is increased. 
If one plots the binodal, or melting, temperature, $T(\rho)$, of the fluid-sold phase boundary as 
a function of the density ($\rho$), or the pressure ($p$), it first steeply increases with the density 
like a standard molecular systems but it reaches a reentrant peak and becomes a decreasing function.   
At high densities beyond this peak, many systems exhibit multiple solid phases, 
each of which is characterized by distinct crystalline
structures~\cite{F.Stillinger1976,S.Prestipino2005,J.Pamies2009,Y.Zhu2011,W.Miller2011}.   
On the other hand, the $Q^\pm$ class systems are characterized by $\vk$ which can take negative values at
finite $k$'s.
A representative example is the generalized exponential model (GEM) defined by 
$v(r) = \epsilon \exp\left[- (r/\sigma)^{n}\right]$ with $n > 2$, where $\epsilon$ and $\sigma$ are the energy
and length scales, respectively. 
In the $Q^\pm$ class systems, the reentrant transition is either suppressed or absent 
and instead the particles can overlap each other and form clusters consisting of multiple particles bonding
together at high
densities~\cite{W.Klein1994,C.Likos2001,B.Mladek2006,C.Likos2007,B.Mladek2007Oct,E.Lascaris2010,F.Sciortino2013}.   
The melting line becomes a monotonic function of the density in the high density limit. 
The density determines the number of particles in a cluster at low temperatures, 
where very rich multiple crystalline phases characterized by different cluster sizes are
observed~\cite{K.Zhang2010,K.Zhang2012Jun}. 
Peculiar dynamics such as fast intercluster ballistic hopping of particles 
has also been reported~\cite{A.Moreno2007,D.Coslovich2011, M.Montes-Saralegui2013}.
The ultrasoft potential systems are not just idealized theoretical models but
they can be seen in many realistic macromolecular systems, such as star polymers and dendrimers~\cite{C.Likos2006} 
and even in some quantum systems such as the vortices of Type 1.5
superconductors~\cite{F.Cinti2010,F.Cinti2014,R.Diaz-Mendez2015,R.Diaz-Mendez2017}.  
Indeed some model macromolecules have been demonstrated to form clusters at high
densities by simulations~\cite{D.Lenz2012,M.Bernabei2013,M.Slimani2014}. 

The ultrasoft potential systems have also been studied in the context of the glass and jamming transitions.
Binary and polydisperse systems of the harmonic and Hertzian potential systems have been pet models to study both jamming and glass 
transitions~\cite{L.Berthier2009Apr,L.Berthier2009Aug,C.OHern2003,*A.Donev2004,*C.OHern2004,M.vanHecke2010}. 
The glass transitions of these systems in the vicinity of the jamming transition point have been 
extensively investigated~\cite{L.Berthier2009Aug,L.Berthier2009Apr,L.Berthier2010Dec}
and the glass transition line $T_g(\rho)$ of this $Q^+$ systems were found to show the reentrant
behaviors at high densities~\cite{L.Berthier2010Dec}. 
Some anomalies of the jamming/glass transitions associated with the large particle overlap at high densities are
also reported~\cite{L.Wang2012,C.Zhao2011}.  
The GCM is another $Q^+$ class glass former. 
In contrast to typical glass formers, it was demonstrated that the monatomic GCM  undergoes
the glass transition at extremely high densities and  its dynamical asymptotic behaviors agree
unprecedentedly well with the mode-coupling theory (MCT)~\cite{A.Ikeda2011Jan,A.Ikeda2011Aug,A.Ikeda2012,D.Coslovich2016}.

\begin{figure}
 \begin{center}
\includegraphics[width=1.0\columnwidth]{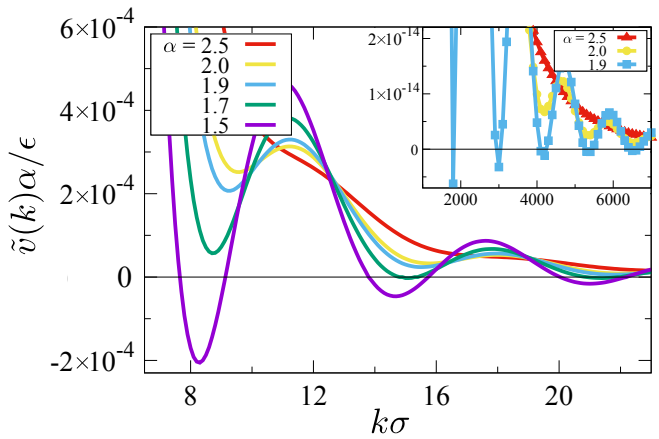}
 \end{center}
 \caption{The Fourier transform $\tilde{v}(k)$ of the generalized Hertzian potential for several $\alpha$'s.
 The inset is $\vk$ for a wider range of $k$'s for $\alpha = 1.9$, 2.0, and 2.5 at large $k$'s.
}
 \label{fig:vk}
\end{figure}

Compared to the $Q^+$ class,  the glass transition of $Q^\pm$ class systems
is largely unexplored~\cite{W.Klein1994,D.Coslovich2012,M.Schmiedeberg2013,R.Miyazaki2016}.
Numerical study of the glass transition has been first reported for GEM with $n=4$, a typical
$Q^\pm$ system, where the particles are clumped to form clusters and these clusters exhibit the glassy slow dynamics at low
temperatures~\cite{D.Coslovich2012}. 
Recently, the cluster glass transition of another type of $Q^\pm$ systems, {\it i.e.}, the
generalized Hertzian potential (GHP) has been reported~\cite{M.Schmiedeberg2013,R.Miyazaki2016}.   
GHP is defined by a pairwise isotropic interaction defined by 
\begin{equation}
v(r) = \left\{
\begin{aligned}
&
\frac{\epsilon}{\alpha} \bigg( 1 - \frac{r}{\sigma} \bigg)^{\alpha}   
\ \ \ \text{for} \ r < \sigma, 
\\
&
0 
\ \ \ \ \ \ \ \ \ \ \ \ \ \ \ \ \  \ 
\text{for} \ r \ge \sigma,
\end{aligned}
\right.
\label{eq:GHP}
\end{equation}
where $\sigma$ is the radius of the particle. 
The harmonic and Hertzian potential correspond to $\alpha=2$ and 2.5, respectively.
By directly Fourier transforming  Eq.~(\ref{eq:GHP}),
one finds that GHP belongs to the $Q^+$ class for $\alpha \geq 2$ 
and to the $Q^\pm$ class otherwise.
Indeed, $\vk$ is analytically written as 
\begin{equation}
\vk = \frac{\epsilon}{\alpha} \frac{4k\sigma - 6\sin(k\sigma) + 2k\sigma \cos(k\sigma)}{(k\sigma)^{5}}.
\label{eq:GHP-FT}
\end{equation}
In Figure \ref{fig:vk}, we plot Eq.~(\ref{eq:GHP-FT}) for several $\alpha$'s.
When $\alpha$ is large, $\vk$ is a monotonically decreasing function and always positive but
if $\alpha$ is decreased below 2, the zeros of $\vk$ start appearing at very large wavevectors. 
The first zero tends to shift to smaller $k$'s as $\alpha$ decreases. 
Therefore, it is expected that the GHP particles with $\alpha < 2$ form clusters at high densities.
In the previous letter~\cite{R.Miyazaki2016}, we have briefly reported that the three dimensional binary GHP
systems with $\alpha <2$ undergo the glass transition and cluster formation 
concomitantly at low temperatures and high densities. 
It was demonstrated that the number of particles per cluster increases one by one
semi-discontinuously as the density increases. 
The phase boundary of the multiple glass phases 
characterized by different cluster sizes was determined and,
surprisingly, the system showed a singular relaxation dynamics in the vicinity of the phase
boundary.

In this paper, we present thorough and detailed analysis of structural, dynamic, and thermodynamic properties
of the GHP cluster glasses for a wide range of densities and for several $\alpha$'s. 
Motivations to study slow dynamics of such an exotic model system are twofold. 
First, the absence of hard-sphere-like excluding volume effect and the presence of extraordinary rich
thermodynamic ground states of the ultrasoft potential systems challenge our conventional and
naive picture of the glass transition that the elementary ingredient of dynamical slow down is the arrest
of the particles inside the rigid cages formed by their neighbors. 
How is this picture modified 
if the particles can penetrate each other?
Secondly, presence of the multiple glass phases observed at high densities 
suggests that the GHP model is an ideal test bench to study the polyamorphism of complex fluids and,
possibly, help understanding the putative liquid-liquid transition underlying at lower temperatures. 
The paper is organized as follows. 
We first introduce the numerical model and explain the method in Section~\ref{sec:model}.
The glass phase diagram, the iso-relaxation-time lines as a function of $T$ and $\rho$, 
for the models with different $\alpha$'s and the static properties are discussed in Section~\ref{sec:Tg}.
We investigate dynamical properties in detail in Section~\ref{sec:dynamics} and we conclude in Section~\ref{sec:conclusion}.

\section{Model and methods}
\label{sec:model}

We investigate a three-dimensional 50:50 binary mixture of large ($L$) and small ($S$) spherical particles
interacting with the GHP potential defined by Eq.~(\ref{eq:GHP}).
For the binary mixture, we replace the radius $\sigma$ with $\sigma_{ab} = (\sigma_{a} + \sigma_{b})/2$, 
where $\sigma_{a}$ is the particle diameter of the $a$ ($\in$ $L$, $S$) component. 
The size ratio of the particles, $\sigma_{L}/\sigma_{S}$, is set to 1.4 and
the mass of the particles is  set to $m$ for both components.
As argued in the previous section, this model belongs to the $Q^\pm$ class when $\alpha <2$. 
There is an argument that if the second derivative of the potential at origin
$v''(r = 0)$ is not negative, the system belongs to the $Q^{\pm}$ class but  this is not the case for GHP 
since $v''(0)$ is always positive even for $\alpha \ge 2$~\cite{C.Likos2007}.

We run the molecular dynamics simulation for the systems with $\alpha = 1.5$, $1.7$, $2.0$, and $2.5$.
We set $\sigma_{S}$, $m$, $10^{-4}\epsilon/k_{B}$ with the Boltzmann constant $k_{B}$, and
$\sqrt{m\sigma_{S}^{2}/\epsilon}$, as the units of the length, mass, temperature, and time,
respectively. 
The simulations are performed in the $NVE$ ensemble with the velocity Verlet
algorithm over a wide range of temperatures $T$ and densities $\rho = N/V$ with
the number $N$ of the particles and the volume $V$ of the simulation box~\cite{D.Frenkel2001}. 
We mainly study the system of $N = 1000$.
We also run simulations for different $N$ at several temperatures and densities and confirmed that no
appreciable finite size effect for our results was observed.
The time step in our simulations is 0.002 for the systems with $\alpha = 1.5$ and $1.7$ and 0.02 for $\alpha = 2$ and $2.5$.
The time step for $\alpha < 2$ is set to be much smaller than that for $\alpha \ge 2$ 
to avoid energy drift in our simulation time windows.
The system was equilibrated at a high temperature ($T = 200$) at first.
We subsequently cooled the system to lower temperatures by normalizing velocities of particles every 5000 steps 
over $10^4$ or $10^5$ time units for the lowest temperatures we investigate.
In order to equilibrate the system after the annealing, every run was performed over 20$\tau$ at least.  
Here $\tau$ is the structural relaxation time defined as the time at which the
intermediate scattering function is equal to 0.1.
We assured that no aging was observed after the equilibration run.

\section{Glass phase diagram and static properties}
\label{sec:Tg}

\subsection{One component systems}
\label{subsec:crystal}

\begin{figure}[tb]
 \begin{center}
	\includegraphics[width=1.0\columnwidth]{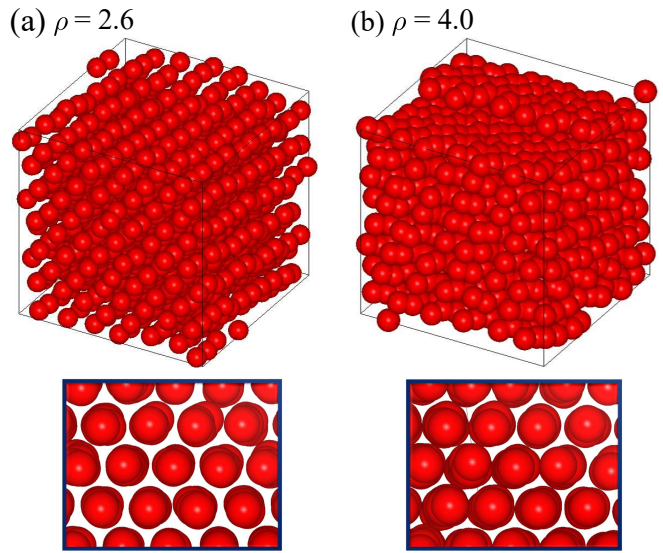} 
 \end{center}
\vspace*{-0.5cm}
 \caption{Snapshots of the one component GHP system with $\alpha=1.5$ for  
(a) $\rho=2.6$, $T=50$ and (b) $\rho=4.0$, $T=50$. 
The particle sizes are not drawn to scale but reduced by 30\%.
Quasi-two dimensional hexagonal phases are observed. 
Particles are seen to align along the columns. The insets are top views 
seen from the direction parallel to the columns.
}
 \label{fig:crystal}
\end{figure}

Before investigating the glass transition of the binary GHP, we briefly summarize our simulation
results for the one component GHP with $\alpha=1.5$ which is well below the $Q^\pm$ criteria of $\alpha=2$.
We run the simulation long enough to assure that the potential energy becomes constant and the system is well
equilibrated. 
The snapshots of the observed crystalline structures are shown in Figure \ref{fig:crystal}.
Contrary to our anticipation, none of the solid phase observed in the simulation boxes is 
the cluster crystals even at densities as high as $\rho=2.6$ and $4.0$. 
Instead, for both densities,  we observe the quasi-hexagonal crystal phase where the particles form the
two dimensional hexagonal layers laminated in the vertical direction of the layers.
This structure is unchanged at higher densities ($\rho=4.0$) but the hexagonal layers are undulated
along the direction parallel to the layers. 
The absence of the cluster phases is consistent with the previous study for the two dimensional
system~\cite{W.Miller2011}, in which series of exotic crystalline phases were obtained over a wide
range of densities but no cluster phase was observed. 
The hexagonal phases which we observed are akin to the lane phase observed in the two dimensional
counterpart~\cite{W.Miller2011}. 
More elaborate free energy assessment using the
thermodynamic integration is necessary in order to determine the accurate and 
full thermodynamic phase diagram of the crystalline structures but it is beyond the scope of this present study. 

\subsection{Glass Phase Diagram}
\label{subsec:Tg}

\begin{figure}
 \begin{center}
\includegraphics[width=1.0\columnwidth]{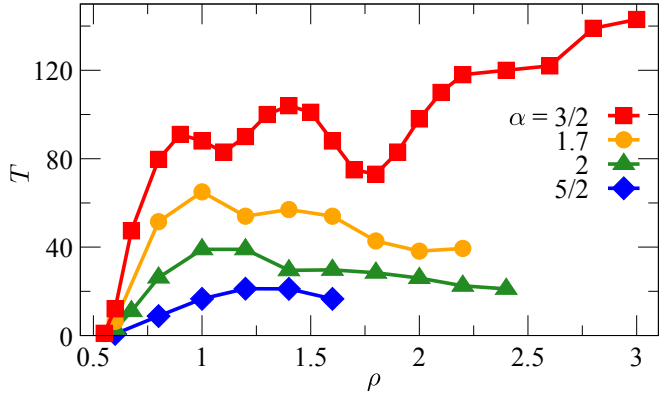}
 \end{center}
\vspace*{-0.5cm}
 \caption{The glass lines $T_g(\rho)$ for $\alpha=1.5$ (square), $1.7$ (circle), $2$ (triangle), and $2.5$ 
(diamond), defined by the temperature at which $\tau$ reaches $10^{3}$.
}
 \label{fig:isotau}
\end{figure}

Let us go back to the binary system and study the glass transition of GHP.
We first draw the glass phase diagram in the $\rho$--$T$ plane for various $\alpha$'s.
We define the fluid-glass phase boundary as the point at which 
the structural (or so-called alpha) relaxation time $\tau$ exceeds a prefixed value. 
$\tau$ is extracted from the intermediate scattering function for the large ($L$) component, 
$\Phi_{L}(k,t)$, defined by 
\begin{equation}
\Phi_{L}(k,t) = 
\dfrac{\lgle \rho_{L}(k,t)\rho_{L}(-k,0)\rgle}{\lgle | \rho_{L}(k,0) |^{2}\rgle}, 
\label{eq:Phikt}
\end{equation}
where 
$\rho_{L}(k,t) = \sum_{j \in \{L\}} \exp [ i \vec{k} \cdot \vec{r}_{j}(t) ]$ 
is the density fluctuations of the large component in the Fourier space,
$\{L\}$ in the summation denotes the set of particles of the large component, 
$\vec{r}_{j}(t)$ is the position of the $j$-th particle. 
We define $\tau$ by $\Phi_{L}(k=6.0,\tau) = 0.1$.
Here, the wavenumber $k = 6.0$ is chosen to be close to the position of the nearest neighbor peak of
the static structure factor $S_{LL}(k)$ for the large component (see Eq.~(\ref{eq:Sk}) below).
As we shall discuss below, $\tau$ for the  correlation functions for the small component or
for their self-part are different at high densities and small $\alpha$'s.  
However, it does not qualitatively affect the overall shape of the phase diagram shown here. 
We defined the empirical glass transition temperature $T_{g}$ as the point at which $\tau$ reaches $10^{3}$.
We run the simulation for a range of densities, $0.5 \le \rho \le 3.0$. 
The glass transition lines thus obtained are shown in Figure~\ref{fig:isotau} for $\alpha = 1.5$, $1.7$, $2.0$, and $2.5$.
They are the iso-$\tau$-lines whose shape can shift up and down depending on the definition of $T_g$
but the overall shapes are unchanged.
For $\alpha = 2.5$, $T_{g}(\rho)$ has the maximum peak around $\rho \approx 1.2$ before it gradually decreases at higher densities.
Above $\rho\approx 1.6$, the system is found to nucleate to form the fcc crystal and no glassy behavior is observed. 
This reentrance behavior is consistent with the Likos criteria for the $Q^+$ class systems.
Similar behavior is reported for the polydisperse system with the same potential~\cite{L.Berthier2010Dec}.   
Asides that the polydisperse system did not crystallizes at high densities, $T_g(\rho)$ agree with 
that for the binary system almost quantitatively, hinting that the result is insensitive to the size dispersity of particles.
The result for $\alpha = 2.0$ is also qualitatively similar; $T_g(\rho)$ increases, reaches the maximal
peak, and then gradually decreases. 
A cusp is observed slightly above the reentrant peak, around $\rho=1.4$, which may be a precursor of
a new glass phase but we did not observe any qualitative structural changes there (not shown).  
We again could not access higher densities above $\rho\approx 2.4$ as demixing and 
nucleation intervene the glassy slow dynamics. 
Interestingly, the position of the peak is close to the phase boundary of the fcc and bcc crystals 
of the one-component counterpart~\cite{Y.Zhu2011}.
Overall shape of $T_g(\rho)$ for $\alpha = 2.5$ and $2.0$, including the positions of the reentrant
peaks, are similar to the fluid-crystal binodal lines for the one-component systems, 
although the latter lie at higher temperatures~\cite{J.Pamies2009,Y.Zhu2011}. 
As $\alpha$ is lowered below 2, the overall trends of $T_g(\rho)$ start changing qualitatively.
For $\alpha = 1.7$, $T_{g}(\rho)$ is enhanced and two minima of $T_{g}(\rho)$ are clearly observed
around $\rho \approx 1.2$ and $2.0$.
For $\alpha = 1.5$, the deeper minima appear at $\rho \approx 1.1$ and $1.8$ and, furthermore, 
$T_g(\rho)$ increases at higher densities, instead of decreasing as observed for systems with 
$\alpha > 2$, which is consistent with the mean-field prediction for the $Q^\pm$ class
fluids~\cite{C.Likos2001} and the results for other cluster-forming systems~\cite{K.Zhang2010,
K.Zhang2012Jun, D.Coslovich2013}.   
The presence of the distinct minima observed for $\alpha <2$ is a clear sign that there exist
multiple glass phases~\cite{R.Miyazaki2016}. 
In the following, we mainly focus on the system with $\alpha=1.5$ 
and discuss thoroughly about static, thermodynamic, and dynamic properties of the system in the
vicinity of $T_g(\rho)$.

\subsection{Static structures}
\label{subsec:statics}

\begin{figure}
 \begin{center}
\includegraphics[width=1.0\columnwidth]{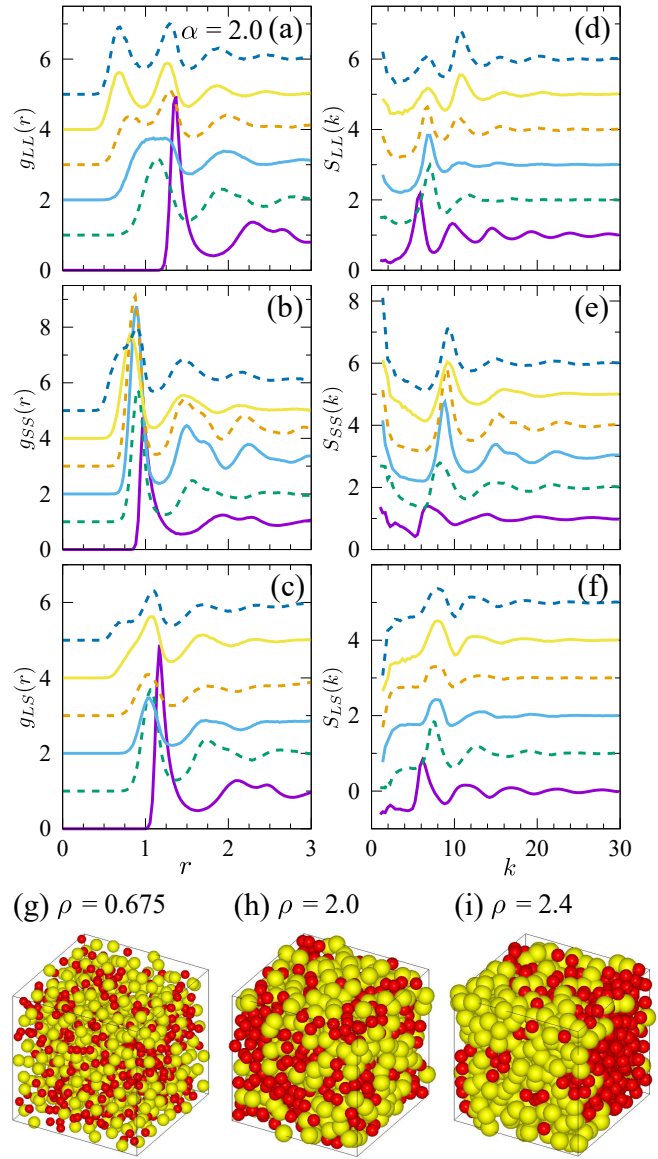}
 \end{center}
 \caption{
(a)--(c) Partial radial distribution functions for $\alpha = 2.0$.
 (a) $g_{LL}(r)$, (b) $g_{SS}(r)$, and (c) $g_{LS}(r)$ at $T_{g}(\rho)$ for $\rho = 0.675$, 1.2, 1.4, 1.6, 2.0, and 2.4 from bottom to top.
 	The alternate solid and dashed lines are shifted vertically by 1 for clarity.
(d)--(f): the corresponding partial static structure factors.
(g)--(i): the snapshots of the equilibrated systems at 
(g) $\rho=0.675$ at $T= 6$, (h) $\rho=2.0$ at $T= 21.5$, and (i) $\rho= 2.4$ at $T=20$.
}
 \label{fig:gofr_Sofk_alpha2.0}
\end{figure}
\begin{figure}
 \begin{center}
  \includegraphics[width=1.0\columnwidth]{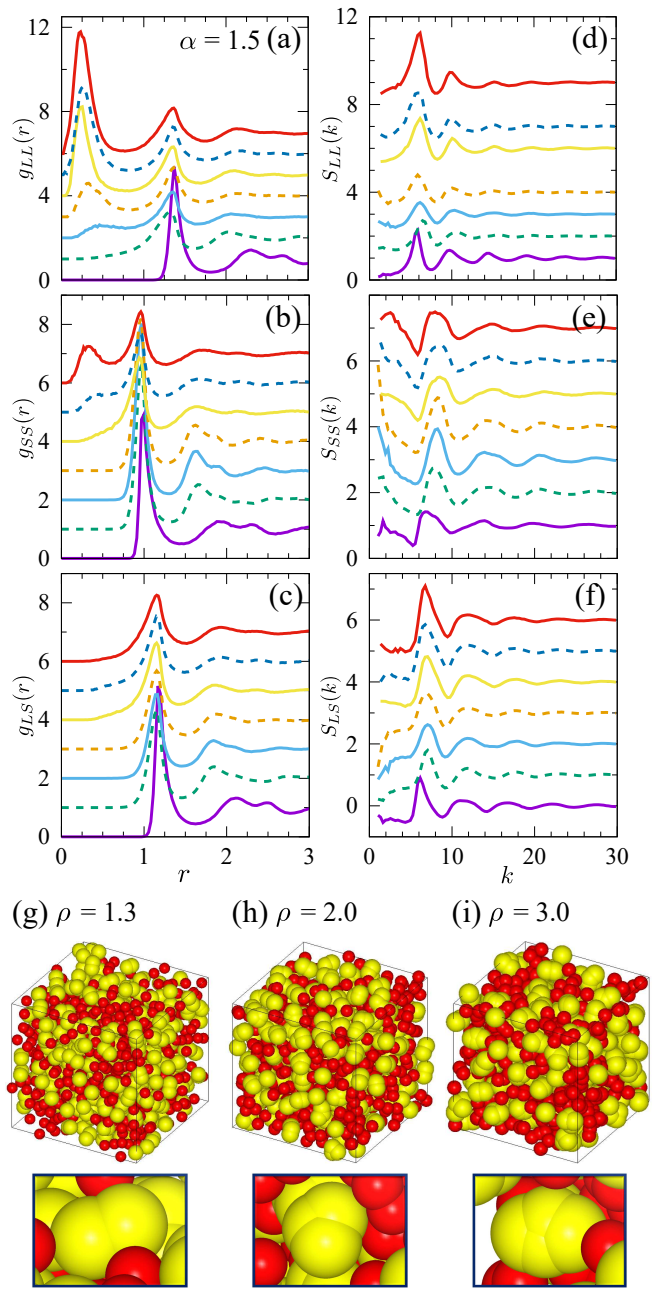}
 \end{center}
 \caption{
(a)--(c) Partial radial distribution functions for $\alpha = 1.5$.
 	(a) $g_{LL}(r)$, (b) $g_{SS}(r)$, and (c) $g_{LS}(r)$ at $T_{g}(\rho)$ for $\rho = 0.675$, 1.0, 1.1, 1.2, 1.7, 1.9, and 2.2 from bottom to top.
 	The alternate solid and dashed lines are shifted vertically by 1 for clarity.
(d)--(f): the corresponding partial static structure factors.
(g)--(i): the snapshots of the equilibrated systems at (e) $\rho=1.3$ at $T = 98$, (f) $\rho=2.0$ at 
$T = 86$, and (g) $\rho=3.0$ at $T = 135$.
 The particle sizes are not drawn to scale but reduced by 30\%.
The insets are their close-ups highlighting dimer-, trimer-, and tetramer-shaped clusters of the large particles.
}
 \label{fig:gofr_Sofk_alpha1.5}
\end{figure}

We investigate static properties of systems with $\alpha=1.5$ 
and compare with those of $\alpha=2.0$ in the vicinity of $T_{g}(\rho)$ shown in Fig.~\ref{fig:isotau}
for several densities.
We evaluate the partial radial distribution functions $g_{ab}(r)$ ($a$, $b \in L$, $S$) defined by
\begin{equation}
g_{ab}(r)= \frac{V}{4\pi r^2 N_a N_b}\sum_{i\in \{a\}}\sum_{j\in \{b\}}\lgle \delta({r} -|\vec{r}_i - \vec{r}_j|)\rgle
\end{equation}
and their Fourier transformations, the partial static structure factors $S_{ab}(k)$ ($a$, $b \in L$, $S$) defined by
\begin{equation}
 S_{ab}(k)= \frac{1}{\sqrt{N_a N_b}}\lgle \rho_a(k)\rho_b^{\ast}(k)\rgle.
 \label{eq:Sk}
\end{equation}
We show $g_{ab}(r)$, $S_{ab}(k)$, and several snapshots of the systems
for $\alpha = 2.0$ in Fig.~\ref{fig:gofr_Sofk_alpha2.0}
and for $\alpha = 1.5$ in Fig.~\ref{fig:gofr_Sofk_alpha1.5}, respectively. 
First look at Fig.~\ref{fig:gofr_Sofk_alpha2.0} for $\alpha=2.0$.  
In Fig.~\ref{fig:gofr_Sofk_alpha2.0}~(a), one observes that,   
when the density is small, 
$g_{LL}(r)$ is qualitatively similar to those of simple liquids and 
characterized by a sharp nearest-neighbor peak at $r \approx \sigma_{ab}$ followed by oscillatory
tail.
As the density increases (see $\rho=1.2$ and 1.4), the nearest-neighbor peak shifts to smaller $r$ and the peak width
broadens, which is a reflection that the particles are squashed and start overlapping. 
At $\rho \geq 1.6$, the nearest-neighbor peak of $g_{LL}(r)$ splits into two. 
The separation between the two peaks gradually increases as the density increases 
and the peak at smaller $r$ moves to and stops at $r \approx 0.7 = \sigma_{L}/2$, whereas 
the peak at larger $r$ moves back to $r\approx \sigma_{L}$ at the highest density, $\rho=2.4$.
Similar behaviors were reported for the same systems with the pressure control simulation
at a finite temperature~\cite{L.Wang2012} and at $T=0$~\cite{C.Zhao2011}.
Note that the position of the small-$r$ peak does not move after it reaches around $\sigma_{L}/2$ at high densities. 
We checked that the peak position does not change appreciably as the temperature is lowered at a
fixed density. 
The split of the peaks of $g_{LL}(r)$ is a consequences of the particles' overlap 
but the fact that the two peaks are located at $r=\sigma_L/2$ and $\sigma_L$ 
suggests that the particles are overlapped with their neighbors but do not form
clusters. If they were clusters, the inter-particle distance corresponding to 
the large $r$ peak position should be separated more than twice of 
the intra-particle distance corresponding to the small $r$ peak position. 
This observation is reflected in $S_{LL}(k)$ in Fig.~\ref{fig:gofr_Sofk_alpha2.0}~(d). 
It exhibits the growth of the second peak, which corresponds to the additional peak at smaller $r$
in $g_{LL}(r)$ at higher densities. 
We also note that the small and large particles of 
the system at the highest density of $\rho=2.4$ partly demix 
as seen in Fig.~\ref{fig:gofr_Sofk_alpha2.0}~(i).  
This is reflected in mild growth of $S_{LL}(k)$ at small $k$'s. 
In this density, we do not observe the glassy slow dynamics anyway. 
Likewise, $g_{SS}(r)$ in Fig.~\ref{fig:gofr_Sofk_alpha2.0}~(b) shows that the position of the peak shift
to lower $r$ and its width widens as the density increases.
However, the trend is weaker than that observed in $g_{LL}(r)$.  
We observe the onset of the split of the first peak only at the highest density of $\rho \approx 2.4$.  
It is interesting that the similar trend is observed for $g_{LS}(r)$ [see Figure~\ref{fig:gofr_Sofk_alpha2.0}~(c)].
$S_{SS}(k)$ and $S_{LS}(k)$ in Fig.~\ref{fig:gofr_Sofk_alpha2.0}~(e) and (f) 
reflect the properties of $g_{SS}(r)$  and $g_{LS}(r)$.
Note that the increase of $S_{SS}(k)$ at $k\rightarrow 0$ is more dramatic than
$S_{LL}(k)$, suggesting that demixing is predominantly driven by small particles.
The existence of the demixing or the phase separation of binary ultrasoft-potential systems at high densities 
was predicted by a mean-field analysis and using the hypernetted-chain
closure~\cite{A.Louis2000,A.Archer2001,A.Archer2002Nov}

Figure \ref{fig:gofr_Sofk_alpha1.5} is the results for $\alpha = 1.5$ at $T_{g}(\rho)$.
At the lowest density of  $\rho = 0.675$, 
all $g_{ab}(r)$ and $S_{ab}(k)$ are similar to those for $\alpha = 2.0$. 
As the density increases, anomalies first appear in $g_{LL}(r)$.  
The nearest-neighbor peak broaden and its tail stretches to smaller $r$. 
Its position barely changes whereas the height substantially lowers.
At $\rho \approx 1.1$, where $T_{g}(\rho)$ has the first minimum [see Figure~\ref{fig:isotau}], 
another peak appears at $r \approx 0.4$.
Increasing the density further, the small-$r$ peak grows 
but its position slightly moves to smaller $r$ and eventually stops at $r \approx 0.2$.
Furthermore, the minimum between the small-$r$ and the nearest-neighbor peaks deepens. 
This is a clear sign of the cluster
formation~\cite{C.Likos2001,B.Mladek2007Oct,E.Lascaris2010,K.Zhang2012Jun,D.Coslovich2012,R.Miyazaki2016}. 
The presence of the small-$r$ peak, or the cluster peak, at $r < \sigma_{L}/2$ and a deep minimum
($g_{LL}(r) \approx 0$) between the two peaks mean  
that a pair of particles is well separated from other pairs.
This is in stark contrast from the case for $\alpha=2.0$ [Fig.~\ref{fig:gofr_Sofk_alpha2.0}~(a)]  
for which the position of the cluster peak is located at larger $r$ ($\approx \sigma_{L}/2$)
and the depth of the minimum between the two peaks remains finite, 
implying that the inter-particles distances of a fraction of particles are shortened 
because they are pressured but particles are not coagulated yet. 
Figure \ref{fig:gofr_Sofk_alpha1.5}~(g) is the snapshot of the system at $\rho=1.3$.
One can clearly observe dimer-shaped clusters of the large particles at $\rho \gtrsim 1.1$.
The position of the cluster peak corresponds to the bond length of a dimer.
As the density increases further, the large particles form the trimers [Fig.~\ref{fig:gofr_Sofk_alpha1.5}~(h)]
and tetramer [Fig.~\ref{fig:gofr_Sofk_alpha1.5}~(i)].
These observations are also reflected in the shape of the static structure factor.
$S_{LL}(k)$'s for $\alpha=2.0$ in Fig.~\ref{fig:gofr_Sofk_alpha2.0}~(d) demonstrates that the second peak at
higher $k$'s continuously grows as a direct consequence that more particles prefer to interacts with
shorter distance, whereas the shape of $S_{LL}(k)$'s for $\alpha=1.5$ [Fig.~\ref{fig:gofr_Sofk_alpha1.5}~(d)]
is qualitatively unchanged, which suggests that clusters formed by particles with the
short bond length are separated by $\sigma_L$ from their neighbors. 

Similar cluster formation is also observed for small particles but at higher
densities as seen in Fig.~\ref{fig:gofr_Sofk_alpha1.5}~(b) and (e). 
The cluster peak of $g_{SS}(r)$ shows up at $\rho \approx 1.9$ which is 
close to the second minimum of $T_{g}(\rho)$ (Fig.~\ref{fig:isotau}).
This peak also grows as the density increases and is located at less than $\sigma_{S}/2$ like for the large particles.
However, its height is lower and the minimum between the cluster peak and the nearest-neighbor peak
is higher than those for the large particles.
These features suggest that 
the the clusters of the small particles are more unstable and the size of the clusters fluctuates.
Contrary to $g_{LL}(r)$ and $g_{SS}(r)$, no cluster peak is observed in $g_{LS}(r)$ for all densities we studied 
as shown in Fig.~\ref{fig:gofr_Sofk_alpha1.5}~(c).
This means that the cluster formation always takes place between the particles of the same size.
The large particles do not pair with the small ones. 
This was observed also in a binary mixture of GEM~\cite{D.Coslovich2012}.
Distinct from the cluster phases of GEM is the shape of the clusters. 
For GEM, particles in one cluster are completely overlapped with each other and the averaged bond length is $0$
with broad fluctuations. This is natural because the shape of the GEM potential is flat at $r=0$, 
so that the particles can sit on top of each other with no energetic penalty.
On the other hand, the potential of GHP with $\alpha <2$ remains repulsive up to $r=0$ and the
particles bond together at a finite distance determined by the characteristic length encoded in $\vk$.

\begin{figure}
 \begin{center}
\includegraphics[width=0.9\columnwidth]{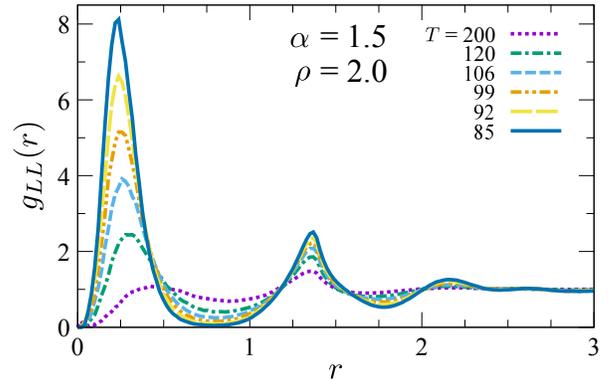}
 \end{center}
 \caption{$g_{LL}(r)$ for $\alpha = 1.5$ at $\rho = 2.0$ at several temperatures.}
 \label{fig:gofr_alpha1.5_rho2.0}
\end{figure}
Finally, in Figure~\ref{fig:gofr_alpha1.5_rho2.0}, the temperature dependence of 
$g_{LL}(r)$ for $\alpha=1.5$ for a fixed density ($\rho=2.0$) is shown. 
At high temperatures $T=200$, $g_{LL}(r)$ show no sharp peak and almost flat even at $r < \sigma_{L}$, 
meaning that the particles can easily penetrate each other.
We can see a precursor of the cluster peak at $r \approx 0.4$.
As the temperature is lowered, both the cluster and nearest neighbor peaks grow and sharpen
but the growth of the cluster peak is more pronounced than the nearest neighbor peak. 
The minimum between the peaks concomitantly decreases.  
This result shows that the cluster formation is crossover and no sharp phase transition 
from the fluid phase to a ``cluster-fluid'' phase exists, which is also the case for the binary mixture of
GEM~\cite{D.Coslovich2012}.

\subsection{Cluster sizes}
\label{subsec:clusters}

\begin{figure}
 \begin{center}
  \includegraphics[width=1.0\columnwidth]{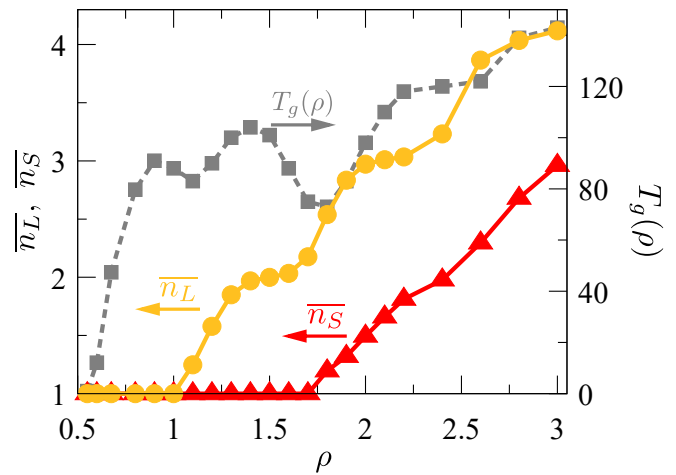}
 \end{center}
 \caption{
The average particle number per cluster of large particles $\overline{n_{L}}$ (circles) 
and small particles $\overline{n_{S}}$ (triangles) and $T_{g}(\rho)$ (squares) for the system with $\alpha = 1.5$.
}
 \label{fig:population}
\end{figure}
We  investigate the properties of clusters formed in the system with $\alpha = 1.5$ in detail. 
As discussed in the previous subsection, the cluster formation is evident from the shape of 
$g_{LL}(r)$ and $g_{SS}(r)$. 
In order to evaluate the number of particles contained in one cluster, one has to integrate
$g_{aa}(r)$ over a cut-off distance $r_{c}$.
We define $r_c$ to be a center of the minimum of $g_{ab}(r)$ between the cluster and nearest
neighbor peaks.
Since all clustering  systems develop clear minima and their positions are almost independent of
temperatures and densities,  we fix  $r_{c, LL}$, $r_{c, SS}$, and $r_{c, LS}$ to be 0.7, 0.6, and 0.65, respectively
and use these values for all densities and temperatures.  
Hereafter, we consider the number of particles per cluster for the large
($n_{L}$) and small ($n_{S}$) particles near $T_{g}(\rho)$. 
The average numbers of $n_{L}$ and $n_{S}$ are shown together with $T_g(\rho)$ in Fig.~\ref{fig:population}.
At low densities, both $n_{L}$ and $n_{S}$ are unity and no cluster is formed.
As the density increases, $n_{L}$ increases by one.  
This stepwise increase takes place at $\rho \approx 1.1$ 
where $T_{g}(\rho)$ shows the first minimum and the cluster peak of $g_{LL}(r)$ starts growing
(Fig.~\ref{fig:gofr_Sofk_alpha1.5} (a)).  
As the density increases further, $n_{L}$ increases one by one and the system enters from the dimer
($n_L=2$) to trimer ($n_L=3$), and to the tetramer ($n_L=4$) phases. 
The density where a stepwise increase is observed synchronizes with the minimum of $T_g(\rho)$. 
On the other hand, clustering of the small particles starts at higher densities.
The dimer phase ($n_S=2$) appears exactly at the density where the large particles enter the
trimer phase and the increase to $n_S=3$ takes place at $\rho \approx 2.6$  where the
large particles becomes tetramers, although the increase of $n_S$  is less sharp.
The multiple cluster phases with relatively sharp phase boundary shown in
Fig.~\ref{fig:population}, however, are already seen at relatively high temperatures. 
In other words, the cluster formation is not triggered by the glass transition nor vice versa. 
\begin{figure}
 \begin{center}
\includegraphics[width=1.0\columnwidth]{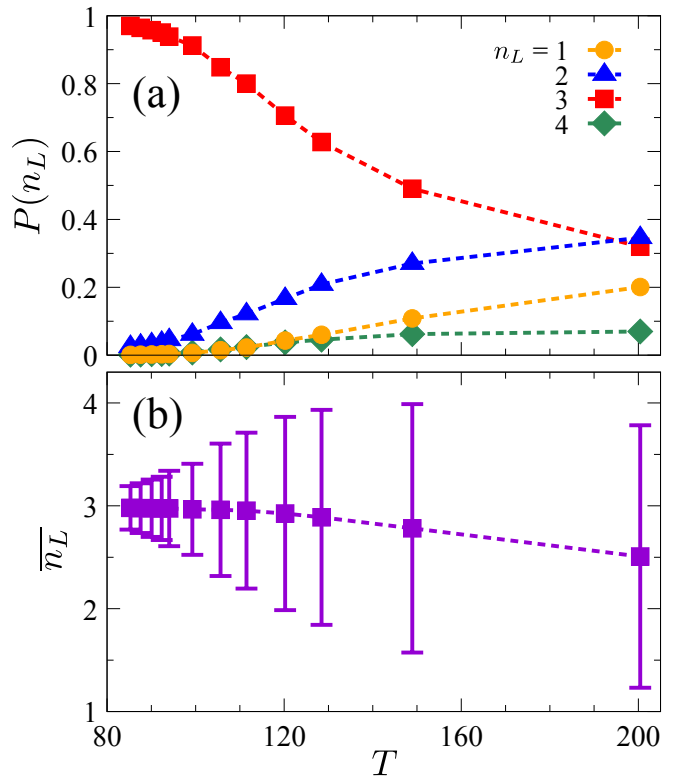}
 \end{center}
\caption{(a) A fraction of large particles participating to clusters with the size of $n_{L}$. 
(b) Temperature dependence of $\overline{n_{L}}$ for the system with $\alpha = 1.5$ at $\rho = 2.0$. 
}
 \label{fig:population_rho2.0}
\end{figure}
In Figure~\ref{fig:population_rho2.0} (a), we show  the temperature dependence of the fraction of the
large particles forming a cluster of the size $n_L$ at $\rho=2.0$ where the large particles are in the trimer phase.  
At very high temperatures, the system consists of the mixture of clusters of various sizes from
$n_L=1$ to $n_L=5$ but as the temperature is lowered, the system is gradually dominated by the
trimer and eventually the system is almost completely occupied by trimers at $T_g(\rho) \approx 98$. 
Temperature dependence of the average of the number of the cluster $\overline{n_{L}}$ also confirm 
that there is no sharp transition from the monomer phase to the trimer phase 
(Fig.~\ref{fig:population_rho2.0}~(b)). 
The gradual but monotonic reduction of the error bars shown in this figure at lower temperatures is a clear sign that
the emergence of $n_L=3$ phase is a crossover. 
Clusters of small particles show similar trend except that the variances are larger than the large
particles (not shown).

\begin{figure}
 \begin{center}
\includegraphics[width=1.0\columnwidth]{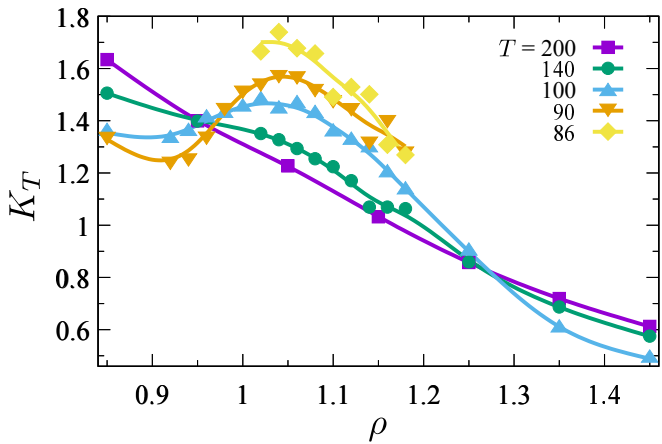}
 \end{center}
 \caption{
 Density dependence of the isothermal compressibility for $\alpha = 1.5$ at several temperatures.
}
 \label{fig:KT}
\end{figure}

The stepwise increase of the cluster size with density, whose edges tend to sharpen 
as the temperature is lowered, hints that the crossover between the different cluster phases 
becomes thermodynamic transition as the temperature is lowered, {\it i.e.},  there might exist 
the liquid-liquid phase transition hidden at a low temperature which is not accessible due to inhibitedly
slow glassy dynamics~\cite{Y.Katayama2000,P.Poole1997,V.Holten2012}.  
This transition may be a liquid version of the first-order transition between the multiple 
crystalline phases which has been reported for the monatomic GEM at high densities and low
temperatures~\cite{K.Zhang2010,K.Zhang2012Jun}. 
As it is often the case of the studies of the polyamorphism and associated liquid-liquid phase
transition, the precursor of the transition can be seen as the Widom line of the isothermal
compressibility at higher temperatures.
We estimate the isothermal compressibility $K_{T}$ defined by
\begin{equation}
 K_{T} = \left\{\rho\left(\pdif{P}{\rho}\right)_{T}\right\}^{-1}
\end{equation}
by evaluating the pressure $P$ for various densities at fixed temperatures around the first minima
of $T_g(\rho)$, where the monomer-dimer phase boundary is observed for the large particles. 
The density dependence of isothermal compressibility $K_{T}$ is shown in Figure~\ref{fig:KT}.
At high temperatures, $K_T$ is a monotonically decreasing function of the density.  
As the temperature is lowered, $K_{T}$ develops a peak at $\rho \approx 1.05$. 
Although the temperatures which we explored are still too high to be conclusive, 
this trend is similar to what has been observed for prospective liquid-liquid phase transitions for
other systems~\cite{V.Vasisht2011,J.Luo2014}.

\begin{figure}
 \begin{center}
	\includegraphics[width=1.0\columnwidth]{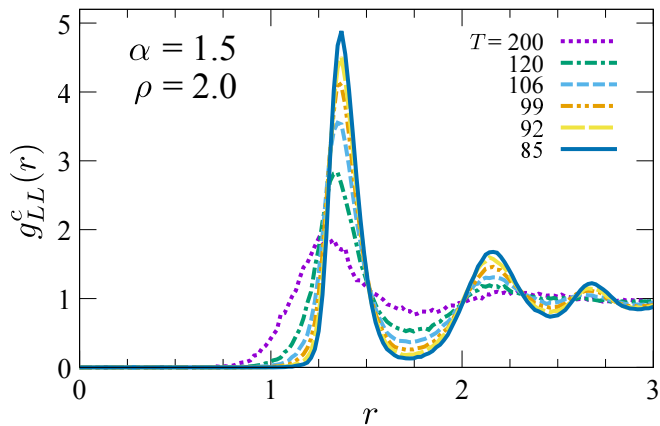}
 \end{center}
 \caption{
	Radial distribution function of the centers of mass of clusters for the large particles 
for $\alpha = 1.5$ at $\rho = 2.0$ and several temperatures.
}
 \label{fig:gofr_cluster_alpha1.5_rho2.0}
\end{figure}

Finally, we assess the static properties of configurations of the clusters.
We compute the partial radial distribution function $g^{c}_{LL}(r)$ of clusters' center of mass.
Figure~\ref{fig:gofr_cluster_alpha1.5_rho2.0} is $g^{c}_{LL}(r)$ for $\alpha = 1.5$ 
in the trimer
phase at $\rho = 2.0$ at several temperatures near $T_g(\rho)$.
It is single-peaked near $r \approx \sigma_L$ which corresponds to the nearest
neighbor distance between the trimers and the peak monotonically increases as the temperature is lowered.
Similar trends are observed at other densities and for $g_{SS}^{c}(r)$.
This means that slow dynamics of the cluster fluids near its glass transition is 
caused by cage formation of clusters much the same way as the cage formation of atoms or molecules 
causes the glass transition of simple liquids. 

\section{Dynamics}
\label{sec:dynamics}

In this section, we discus slow dynamics near $T_g(\rho)$ of clustered fluids 
by analyzing the intermediate scattering functions.
We specifically focus on the collective and self part of the scattering functions for the same species. 
The collective part is written explicitly as 
\begin{equation}
\Phi_{a}(k,t) = 
\frac{1}{S_{aa}(k)}\lgle \frac{1}{N_a} \sum_{i,j \in \{a\}} e^{i \vec{k} \cdot [\vec{r}_{i}(t) - \vec{r}_{j}(0)] } \rgle
\label{eq:Phi2}
\end{equation}
and the self part is
\begin{equation}
\Phi_{s, a}(k,t) = \lgle \frac{1}{N_a} \sum_{j \in \{a\}} e^{i \vec{k} \cdot [\vec{r}_{j}(t) - \vec{r}_{j}(0)] } \rgle.
\label{eq:Phiself}
\end{equation}

\subsection{Density dependence}
\label{subsec:alldynamics}

We first display $\Phi_{L}(k,t)$, $\Phi_{s,L}(k,t)$, $\Phi_{S}(k,t)$, and $\Phi_{s,S}(k,t)$ for a
wide range of densities for $\alpha=1.5$ for which the rich cluster phases are observed. 
\begin{figure*}
 \begin{center}
  \includegraphics[width=2.0\columnwidth]{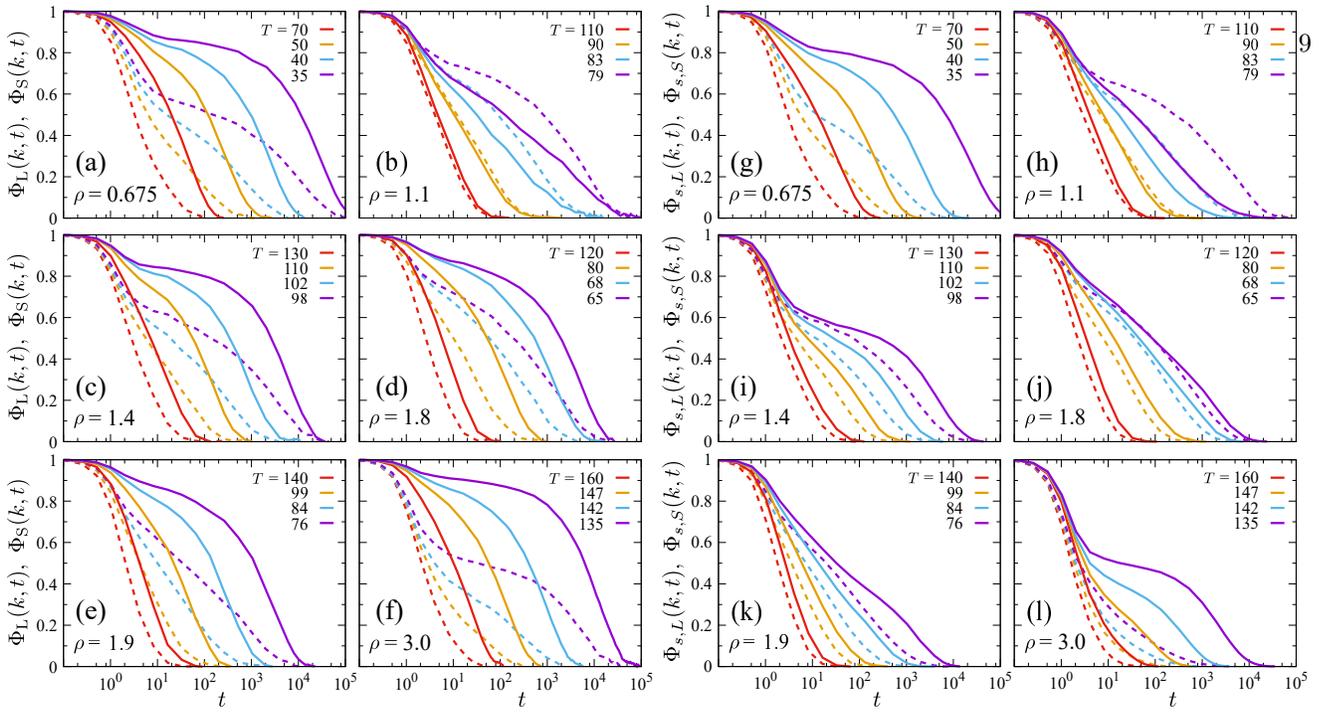} 
 \caption{
 The intermediate scattering functions for $\alpha = 1.5$ and $\rho = 0.675$, 1.1, 1.4, 1.8, 1.9, and 3.0 
 at several temperatures (see legend).
 (a)--(f) The collective parts for the large and small particles, $\Phi_{L}(k,t)$ and $\Phi_{S}(k,t)$, 
 and (g)--(l) the self parts, $\Phi_{s,L}(k,t)$ and $\Phi_{s,S}(k,t)$.
 The solid and dashed lines are for the large and small particles, respectively.
 The wave number $k$ is chosen to be 6.0 for the large and 8.0 for the small particles, respectively.
}
 \label{fig:F_Fs}
 \end{center}
\end{figure*}
Figure~\ref{fig:F_Fs} (a)--(f) are the collective parts for the large and small particles, 
$\Phi_{L}(k,t)$ and $\Phi_{S}(k,t)$ and (g)--(l) are the self parts, $\Phi_{s,L}(k,t)$ and $\Phi_{s,S}(k,t)$ 
for various temperatures and for selected densities,
$\rho = 0.675$, 1.1, 1.4, 1.8, 1.9, and 3.0.
For all figures, $k$ is chosen to be $k = 6.0$ for large and $k= 8.0$ for small
particles, which are close to the first peak of $S_{aa}(k)$ [see Fig.~\ref{fig:gofr_Sofk_alpha1.5}]. 
At low densities, $\rho=0.675$, around which $T_{g}(\rho)$ increases with density, 
both collective and self correlation functions for large and small particles 
exhibit typical slow dynamics of canonical supercooled fluids, {\it i.e.}, the two-step relaxation
characterized by a well-developed plateau followed by the stretched exponential 
relaxation as shown in Figs.~\ref{fig:F_Fs} (a) and (g).
Furthermore, the time-temperature superposition
holds as for standard glass formers (not shown)~\cite{W.Gotze2009}. 
At $\rho =1.1$, 
where $T_g(\rho)$ has a minimum and the large particles start clustering, 
the plateaus of $\Phi_{L}(k,t)$ and  $\Phi_{s,L}(k,t)$ shown in Figs.~\ref{fig:F_Fs} (b) and (h)
disappear and the relaxation qualitatively changes. 
Almost linear decays in the semi-log plot in time means that the relaxation is logarithmic.
The detail of this anomalous relaxation is discussed in the next subsection.
On the other hand, $\Phi_{S}(k,t)$ and $\Phi_{s,S}(k,t)$ for small particles 
still show two-step relaxations. 
At $\rho=1.4$, we find that both $\Phi_{L}(k,t)$ and $\Phi_{s,L}(k,t)$ retrieve the two-step relaxation as
shown in Figs.~\ref{fig:F_Fs} (c) and (i). 
At $\rho=1.9$, where $T_g(\rho)$ shows the second minimum and 
the small particles also start clustering to form dimers,
$\Phi_{L}(k,t)$ remains qualitatively unchanged, but the dynamics of $\Phi_{S}(k,t)$  and $\Phi_{s,S}(k,t)$ 
become singular, showing the logarithmic decay instead of two-step relaxation. 
Furthermore, $\Phi_{s,L}(k,t)$ also shows the logarithmic decay. 
This suggests a decoupling of
the collective and self dynamics at this density for the large particles [see Figs.~\ref{fig:F_Fs} (e) and (k)].
At the highest density $\rho=3.0$, one observes that both
$\Phi_{L}(k,t)$ and $\Phi_{S}(k,t)$ retrieves the two-step relaxation again but two distinct changes
are observed for their self parts, $\Phi_{s,L}(k,t)$ and $\Phi_{s,S}(k,t)$.
First, the self part for the large particles $\Phi_{s,L}(k,t)$ 
retrieves the two-step relaxation but its relaxation time is substantially
shorter than its collective part. 
Secondly, the self part for the small particles 
$\Phi_{s,S}(k,t)$ does not show the two-step relaxation nor logarithmic relaxation and 
the relaxation is much faster than other correlation functions. 
This facilitated relaxation of small particles and decoupling both from its collective dynamics
and those of large particles are reminiscent of the dynamical anomaly observed in the binary mixture of the particles with
disparate size ratio~\cite{Moreno2006h,Moreno2006g}.
This qualitative change from the two-step relaxation to the single-step relaxation is also similar
to the transition from the so-called Type B to Type A dynamics.
Type A and Type B dynamics are the two distinct glassy slow dynamics 
originally predicted by the mode-coupling theory (MCT)~\cite{gotze1992}.
The former refers to a single step relaxation with continuous growth of the height of $\Phi_{a}(k, \tau)$
whereas the latter refers to a two-step relaxation with discontinuous jump of the plateau height of
$\Phi_{a}(k, \tau)$ near the glass transition temperature.
Mixed dynamics of type A and B is often observed when two competing arresting mechanisms, 
such as gelation and glass transitions, are at play~\cite{Zaccarelli2005c,*Zaccarelli2006}. 

\begin{figure}
 \begin{center}
  \includegraphics[width=1.0\columnwidth]{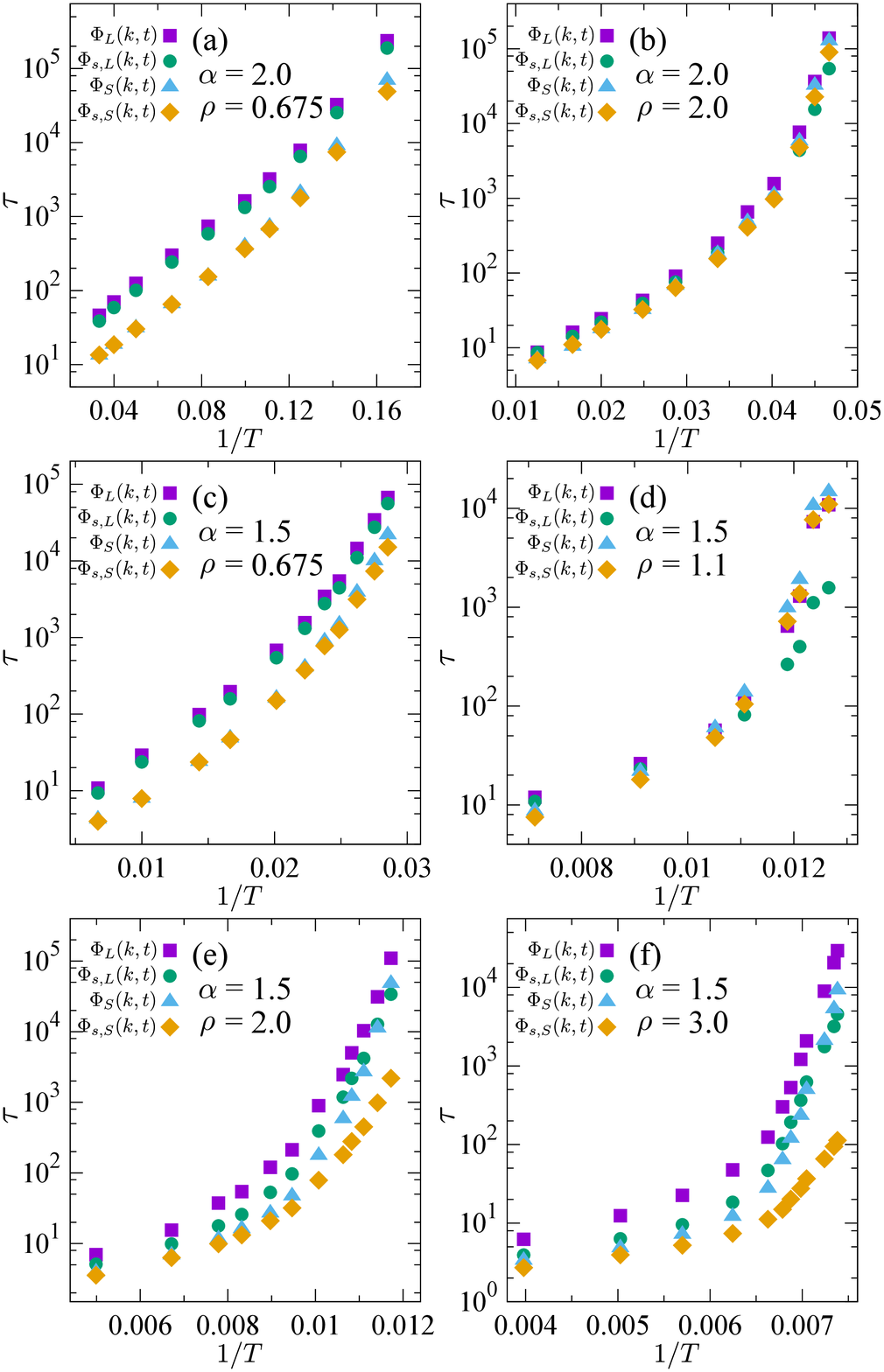}
 \end{center}
 \caption{
	$\tau$ versus $T^{-1}$ for the collective part of the large (square) and the small (triangle) particles 
	and for the self part of the large (circle) and the small (diamond) particles
	for $\alpha = 2.0$ [(a) and (b)] and $\alpha = 1.5$ [(c)--(f)] at densities of 
        $\rho=$0.675 [(a) and (c)], 1.1 [(d)], 2.0 [(b) and (e)], and 3.0 [(f)].
	}
 \label{fig:F_alpha1.5_rho3.0}
\end{figure}
Decoupling of the self and collective relaxation and the large and small particles 
can be lucidly seen in the temperature dependence of the relaxation time $\tau$. 
Figure~\ref{fig:F_alpha1.5_rho3.0} shows the temperature dependence of 
$\tau$ for $\rho=0.675$ and $2.0$ for the non-clustering system with $\alpha=2.0$ (Fig.~\ref{fig:F_alpha1.5_rho3.0}~(a)--(b))
and for several densities for the clustering system with $\alpha=1.5$
(Figs.~\ref{fig:F_alpha1.5_rho3.0}~(c)--(f)). 
All $\tau$'s for $\Phi_L(k,t)$, $\Phi_S(k,t)$, $\Phi_{s,L}(k,t)$, and $\Phi_{s,S}(k,t)$ are plotted. 
It is well established that all relaxation times for both collective and self and for different
components behave similarly for conventional glass models such as the Lennard-Jones and hard sphere
mixtures~\cite{W.Gotze2009}.
This is indeed the case for $\alpha=2.0$ as shown in Fig.~\ref{fig:F_alpha1.5_rho3.0}~(a). 
Aside that $\tau$ for the small particles is slightly smaller, their temperature dependence are
identical. 
Agreement is even better at high densities  for $\rho=2.0$ (Fig.~\ref{fig:F_alpha1.5_rho3.0}~(b)).
For $\alpha=1.5$, $\tau$ at the low density $\rho=0.675$ is almost identical with
those for $\alpha=2.0$ (Fig.~\ref{fig:F_alpha1.5_rho3.0}~(c)).
Interestingly, even at the density $\rho=1.1$ where the anomalous logarithmic decay is observed for $\Phi_L(k,t)$ and
$\Phi_{s,L}(k,t)$, $\tau$'s for all correlation functions are similar except for small
deviation of $\tau$ for $\Phi_{s,L}(k,t)$ at the lowest temperature.
The situation changes at higher densities. 
At $\rho=2.0$, the four $\tau$'s deviate from each other
(Fig.~\ref{fig:F_alpha1.5_rho3.0}~(e)).
$\tau$'s for the collective parts tend to be larger than those of the self parts. 
Especially, $\tau$ for $\Phi_{s,S}(k,t)$ is substantially smaller and its temperature dependence
is weaker than other $\tau$'s.
This trend is more enhanced at higher density, $\rho=3.0$ 
(Fig.~\ref{fig:F_alpha1.5_rho3.0}~(f)).
These observations are consistent with overall trend demonstrated in Fig.~\ref{fig:F_Fs} and strongly indicates 
decoupling of dynamics between the self and collective fluctuations 
and between small and large particles.
Dynamics of the self part of small particles is the fastest and its relaxation is very different from the
typical glassy dynamics.

\subsection{Relaxations of the center of mass correlations}
\label{subsec:COM}

\begin{figure}
 \begin{center}
\includegraphics[width=0.9\columnwidth]{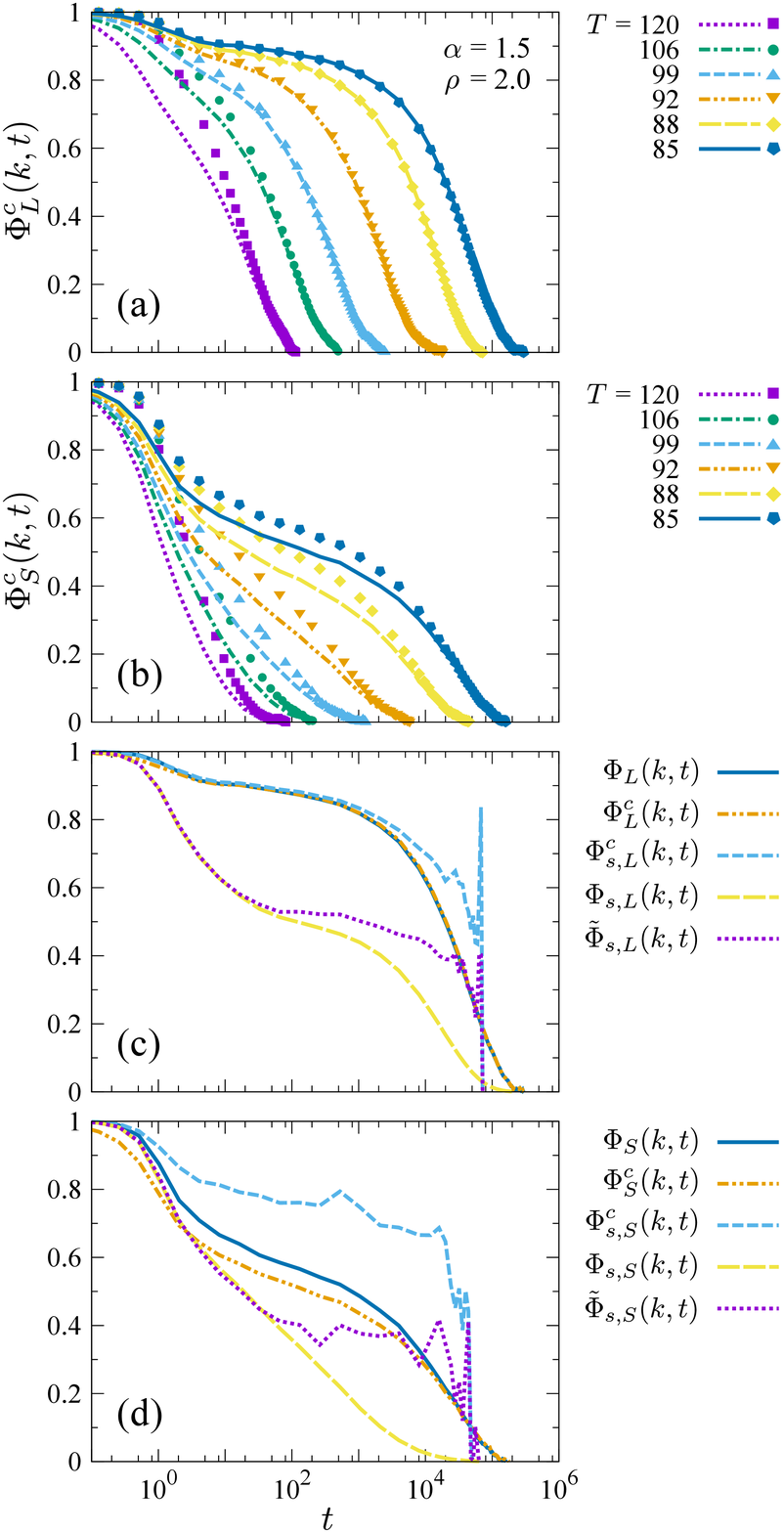}
 \end{center}
 \caption{
The intermediate scattering functions defined in several ways (see text)
 for the system with $\alpha = 1.5$ at $\rho = 2.0$ and several temperatures.
 	(a) The lines is $\PhicL$ for large particles defined by Eq.~(\ref{eq:Phic}) and 
 the dots is $\PhiL$.
 	(b) The corresponding functions for the small clusters and particles.
 	The several intermediate scattering functions of the (c) large and (d) small particles 
 	for the system at $T = 85$.
	$\PhicL$ and $\tPhisL$ are defined by Eqs.~(\ref{eq:Phic}) and (\ref{eq:tPhis}).
 $\PhicsL$ is the corresponding self part defined in the same way as Eq.~(\ref{eq:Phic}).
}
 \label{fig:Fc_alpha1.5_rho2.0}
\end{figure}

We investigate microscopic origins of differences of dynamics between the collective and self
 correlations and between the small and large particles.
Similar decoupling of glassy dynamics has been also reported for a binary mixture of GEM near the glass transition
temperature~\cite{D.Coslovich2012}, according to which the faster relaxation of 
the self part is due to intercluster hopping of particles.
The facilitation of the relaxation due to hopping is even more spectacular for the cluster crystals
of a monatomic GEM system, where the particles hop over 10 nearest neighbors distances in a single
hopping event~\cite{A.Moreno2007,D.Coslovich2011,M.Montes-Saralegui2013}. 
We first introduce the collective intermediate scattering function for the clusters, 
defined by the center of mass of clusters instead of particles as
\begin{equation}
\Phi^{c}_{a}(k,t) = \lgle \frac{\rho^{c}_{a}(k,t)\rho^{c}_{a}(-k,0)}{| \rho^{c}_{a}(k,0) |^{2}} \rgle, 
\label{eq:Phic}
\end{equation}
where $\rho^{c}_{a}(k,t) = \sum_{j \in \Omega^{c}_{a}} \exp [i \vec{k} \cdot \vec{r}^{c}_{j}(t)]$.
$\Omega^{c}_{a}$ denotes the set of indices for the clusters of the component $a$ 
and $\vec{r}^{c}_{j}$ is the position of the center of mass of the $j$-th cluster.
The self part $\Phi^{c}_{s,a}(k,t)$ is defined in the same way.
Note that the self part is calculated only during the time interval in which a particle
resides in a single cluster. 
Figure~\ref{fig:Fc_alpha1.5_rho2.0} shows $\Phi^{c}_{L}(k,t)$ and $\Phi^{c}_{S}(k,t)$ 
for $\alpha=1.5$ at $\rho = 2.0$, 
at which the system is in the trimer phase. 
$\Phi_{L}(k,t)$ and $\Phi_{S}(k,t)$ are also shown in the figure 
for comparison.
At high temperatures ($100 \lesssim T \lesssim 120$), 
$\Phi^{c}_{L}(k,t)$ decays faster than $\Phi_{L}(k,t)$ in short times, 
while it merges to $\Phi_{L}(k,t)$ at long times.
This faster relaxation can be understood as the instability of clusters at the high temperatures
as demonstrated in Figs.~\ref{fig:gofr_alpha1.5_rho2.0} and \ref{fig:population_rho2.0}, where 
$g_{LL}(r)$ exhibits a low and broad peak. 
The positions of the center of mass of the clusters are easily changed by annihilation or
rearrangement by thermal fluctuations.
At long time, however, $\Phi^{c}_{L}(k,t)$ almost collapses to $\Phi_{L}(k,t)$ for all
temperatures. 
This result eloquently demonstrates that the collective slow dynamics of $\Phi_{L}(k,t)$ is 
governed by slow dynamics of clusters each of which behaves like a rigid molecule. 
Similar behavior is observed for the  small particles. 
However, $\Phi^{c}_{S}(k,t)$  decays slightly faster than $\Phi_{S}(k,t)$.  
This is speculated as the consequences that the clusters of small particles are more unstable than
the larger ones as evidenced as a lower and broader peak of $g_{SS}(r)$ 
[see Figs.~\ref{fig:gofr_alpha1.5_rho2.0} and \ref{fig:population_rho2.0}].
From these observations, we conclude that glassy slow dynamics of both large and small collective
fluctuations are governed not by individual particles but by the clusters. 

Next, let us compare the self and collective dynamics. 
Substantial difference of the shapes of the collective and self correlation functions
and their relaxation times discussed in the previous subsection suggests  that
there exists another relaxation mechanism in the self-correlation functions to facilitate dynamics. 
To clarify this point, we introduce 
another function called the single particle correlation function {\it without hopping} defined by 
\begin{equation}
\widetilde{\Phi}_{s, a}(k,t) = \lgle \frac{1}{\tilde{N}_{a}(t)} \sum_{j \in \tilde{a}(t)} e^{i \bm{k} \cdot [\bm{r}_{j}(t) - \bm{r}_{j}(0)] } \rgle, \label{eq:tPhis}
\end{equation}
where $\tilde{a}(t)$ denotes a set of particles of the $a$-th component which 
belong to the same clusters in a time window $(0, t)$.
We expect this function to detect particles which {\it have not} left the
cluster during this time window. 
This function is introduced in order to assess  the contribution of the intercluster particle hopping
during the relaxation. 
In Figure~\ref{fig:Fc_alpha1.5_rho2.0}, we show the results 
at a large density and at a low temperature,
where the characteristics of the two functions are most clearly seen.
We observe several interesting features, in particular, for the large particles.
First, the self-part of the center-of-mass correlator $\PhicsL$ agrees very well with 
$\PhiL$ and $\PhicL$ from the short to intermediate time windows where the plateau is well developed,  
before it departs from the plateau. 
Relaxation of $\PhicsL$ is slower than those of
$\PhiL$ and $\PhicL$. 
Eventually $\PhicsL$ abruptly relaxes to zero with jerky oscillations, which is due to 
dissociation and annihilation of clusters at long times. 
This result demonstrates that the self dynamics and the collective dynamics of {\it clusters} are
similar as it is the case for canonical simple glass formers.
In other words, the difference between the collective $\Phi_{L}(k, t)$ and the self
$\Phi_{s,L}(k,t)$ is attributed to the dynamics of the individual particles inside each clusters
and the inter-cluster hopping of the particles.
The effects of the intra-cluster dynamics and the inter-cluster hopping are distinguished 
by comparing $\PhisL$ and $\tPhisL$ in Fig.~\ref{fig:Fc_alpha1.5_rho2.0}~(c). 
The two functions display the same behavior in the $\beta$-relaxation regime, 
while $\PhisL$ shows clearer plateau than $\tPhisL$ and its slow relaxation is similar to $\PhiL$.
This fact demonstrates that 
the intra-cluster dynamics, detected by both functions, 
causes the low plateau height 
and is the primary reason for the different shape of $\PhiL$ in the short-time window. 
Fig.~\ref{fig:Fc_alpha1.5_rho2.0}~(c) also 
shows that the hopping is responsible for the faster relaxation of $\PhisL$ than $\PhiL$.
On the other hand, 
the dynamics of the small particles shown in Fig.~\ref{fig:Fc_alpha1.5_rho2.0}~(d) 
is more complicated than that of the large ones.
The single-particle correlation function without hopping $\tPhisS$ has a higher plateau than $\PhisS$
and the self part of the center-of-mass correlator $\PhicsS$ also displays a higher plateau 
than the collective $\PhicS$ and its single-particle counterpart $\PhiS$.
Since both $\PhicsS$ and $\tPhisS$ do not contain the particles which hop to other clusters, 
one concludes that the inter-cluster exchange of the particles is 
one of the reasons for the faster relaxation of $\PhisS$.
However, the departure of $\PhicsS$ from $\PhiS$ and $\PhicS$ starts before that of $\tPhisS$ from $\PhisS$.
This hints that a different type of dynamics may also exist and affect the relaxation of small
particles.

\subsection{Higher order singularities}
\label{subsec:A4}

\begin{figure}
 \begin{center}
	\includegraphics[width=1.0\columnwidth]{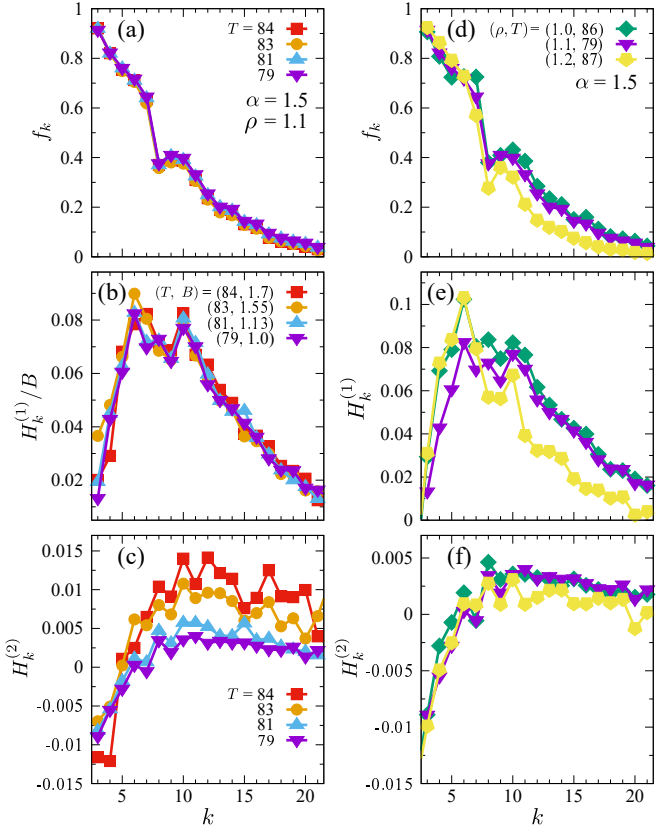}
 \end{center}
 \caption{
  (a) $f_{k}$, (b) $H_{k}^{(1)}/B$, and (c) $H_{k}^{(2)}$ for the system with $\alpha = 1.5$ at $\rho = 1.1$ and 
selected temperatures (see legend), 	
  obtained by fitting $\Phi_{L}(k,t)$ to the asymptotic logarithmic law [Eq.~(\ref{eq:HOS})].
  $t_{0}$ is determined to be 7.
 (d)--(f) $f_{k}$, $\Hk{1}$, and $\Hk{2}$ for $(\rho, T) = (1.0, 86)$, $(1.1, 79)$, and $(1.2,
 87)$. 
}
 \label{fig:HOS_F_alpha1.5}
\end{figure}

\begin{figure}
 \begin{center}
	\includegraphics[width=1.0\columnwidth]{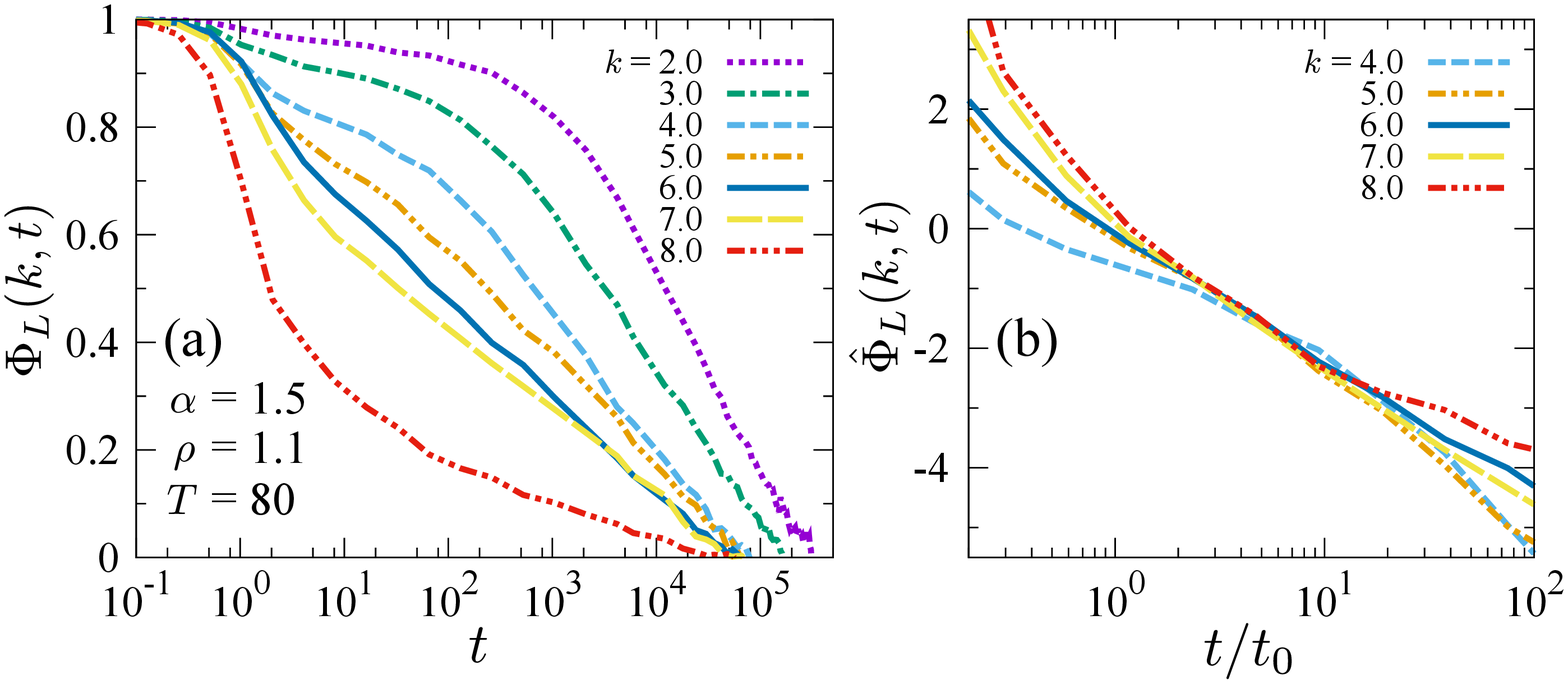}
 \end{center}
 \caption{
 	(a) $\PhiL$ of the system with $\alpha = 1.5$ at $\rho = 1.1$ and $T = 80$ for selected wave numbers (see legend).
 	(b) The rescaled function $\hat{\Phi}_{L}(k,t)$ defined by Eq.~(\ref{eq:Phihat}) with $t_{0} = 7$ for wave numbers around the value for which $H_{k}^{(2)} \approx 0$.
}
 \label{fig:F_alpha1.5_rho1.1_T80}
\end{figure}

The most outstanding dynamical property of the GHP glass is the logarithmic relaxation 
appearing at densities where $T_{g}(\rho)$ exhibits minima and the particles start forming clusters.
As shown in Fig.~\ref{fig:F_Fs}~(b),
this anomalous relaxation is most clearly seen at $\rho=1.1$ where the large particles start forming the dimer.  
The plateau of $\Phi_{L}(k,t)$ disappears and the relaxation becomes logarithmic, 
whereas $\Phi_{S}(k,t)$ still show the two-step relaxation.
The logarithmic relaxation has been originally predicted by 
MCT~\cite{L.Fabbian1999May,L.Fabbian1999Aug,K.Dawson2000} as a signal of the so-called higher-order
($\text{A}_{3}$ or $\text{A}_{4}$) singularity around the end point of the MCT (dynamical) transition
line~\cite{W.Gotze2002, W.Gotze2009}. 
Subsequent simulations and experimental studies demonstrated 
the logarithmic relaxation in various systems, such as 
short-ranged attractive colloids~\cite{F.Mallamace2000,W.Chen2002,S.Chen2003,K.Pham2002,A.Puertas2002,E.Zaccarelli2002Oct,F.Sciortino2003,H.Cang2003Feb,H.Cang2003May}, 
star polymers~\cite{C.Mayer2009}, 
soft colloidal particles~\cite{A.Kandar2012}, 
square-shoulder potential fluids~\cite{G.Das2013, N.Gnan2014}, 
polymer blends~\cite{A.Moreno2006May}, 
binary mixtures of particles with disparate size ratio~\cite{A.Moreno2006Aug,A.Moreno2006Oct, T.Sentjabrskaja2016}, 
proteins~\cite{M.Lagi2009, X.Chu2010}, 
and tRNA~\cite{X.Chu2013}.
This higher-order singularity tends to appear when the two competing glassy dynamics coexist in the
system. 
For the attractive glasses, the glass transition caused by hard-sphere-like repulsion 
competes with the glass transition caused by bonding of particles due to sticky short-ranged attraction.
For binary systems with the disparate size ratio, the two glassy dynamics characterized by disparate
time-scales competes. 
In our system, the coexistence of the two glass phases, the monomer glass and dimer glass 
is expected to be the origin of the logarithmic singularity. 
At the lowest temperature in our system the logarithmic relaxation persisted over four decades and  
this is observed for both collective and self parts of the correlation functions.
To verify that the observed behavior is genuinely due to the higher-order singularity, 
we fit the data for $\Phi_{L}(k,t)$ with the MCT asymptotic function;
\begin{equation}
\Phi_{L}(k,t) \sim f_{k} - H_{k}^{(1)} \ln (t/t_{0}) + H_{k}^{(2)} \ln^{2} (t/t_{0}), \label{eq:HOS}
\end{equation}
where $f_{k}$, $H_{k}^{(1)}$, and $H_{k}^{(2)}$ are the critical non-ergodicity parameters, 
the critical amplitudes of the first, and second orders, respectively~\cite{W.Gotze2009,W.Gotze2002}.
The time scale $t_{0}$ is determined by the plateau height.
We here neglect higher order terms, $\ln^{3}(t/t_{0})$ and $\ln^{4}(t/t_{0})$, 
which can be the same order of magnitude as the term $\ln^{2}(t/t_{0})$ but is predicted to 
vanish in the case of the A$_{4}$ singularity~\cite{W.Gotze2009, W.Gotze2002}.
$f_{k}$, $H_{k}^{(1)}$, and $H_{k}^{(2)}$ calculated for $\rho = 1.1$ are plotted as functions of
$k$ in Figs.~\ref{fig:HOS_F_alpha1.5}~(a)--(c). 
We determined $t_{0}$ in such a way that $f_{k = 6.0} \simeq 0.7$.
The curves of $f_{k}$ for different temperatures collapse on a single function as shown in Fig.~\ref{fig:HOS_F_alpha1.5}~(a).
In Figure~\ref{fig:HOS_F_alpha1.5}~(b), we plot $H_{k}^{(1)}$ 
scaled with a $k$-independent and $T$-dependent function $B(T)$.
This result demonstrates that $H_{k}^{(1)}$ is self-similar in $k$ and can be decomposed as $H_{k}^{(1)} = h_{k}B(T)$.
This is consistent with the prediction of MCT which claims that $B(T)$ is independent of $k$~\cite{W.Gotze2002}.
We also find that $B(T)$ is moderately decreasing function of the temperature.
Figure~\ref{fig:HOS_F_alpha1.5}~(c) shows $H_{k}^{(2)}$.
We find that $H_{k}^{(2)}$ is negative for small $k$ and positive for large $k$, 
and it vanishes at $k \approx 6.0$ which is close to the first peak of the static structure factor for the large particles.
The concave-to-convex crossover as $k$ increases shown in Fig.~\ref{fig:F_alpha1.5_rho1.1_T80}~(a)
is another typical feature of the singularity.
In Figure~\ref{fig:F_alpha1.5_rho1.1_T80}~(b), we 
plot a rescaled function defined by
\begin{equation}
\hat{\Phi}_{L}(k,t) = \frac{ \Phi_{L}(k,t) - f_{k} }{ H_{k}^{(1)} }. \label{eq:Phihat}
\end{equation}
If $\Phi_{L}(k,t)$ is purely logarithmic, $\hat{\Phi}_{L}(k,t)$ is a straight linear function of $\ln (t/t_{0})$.
The deviation from a logarithmic function is found at small and large $k$.
Although our result shares similarity with other systems known to exhibit the singular dynamics,
the wavevector at which $H_{k}^{(2)} \approx 0$ is much larger than those of other 
systems~\cite{E.Zaccarelli2002Oct, F.Sciortino2003, G.Das2013, N.Gnan2014}. 
In our systems, this wavevector is close to $k \approx 2\pi/\sigma_{L}$, 
implying that the microscopic length scale is at play in our model 
and that the mechanism behind the singularity is different from other systems. 
Another difference is that the singular dynamics is most clearly observed 
for the large particles. For other systems, such as the binary mixture with disparate size ratio,
the singular dynamics is observed for small particles~\cite{A.Moreno2006Aug,A.Moreno2006Oct}.

In Figures~\ref{fig:HOS_F_alpha1.5} (d)--(f), the density dependence of 
$f_{k}$, $H_{k}^{(1)}$, and $H_{k}^{(2)}$ is shown. 
One finds no qualitative difference of these parameters slightly below ($\rho = 1.0$)
and above ($\rho= 1.2$) the critical point,  
although the parameters for $\rho = 1.2$ are slightly smaller.
This is in stark contrast with the case of the short-range attractive colloids
for which $f_k$ discontinuously changes in the vicinity of the singular point~\cite{A.Puertas2002,E.Zaccarelli2002Oct}.
Note, however, that the latter is a generic feature of the dynamical singularity but 
it is due to the intervention of the attractive glass transition caused by strong bonds of the colloidal particles.
According to the original formulation of MCT, the higher order dynamical
singularity is not necessarily accompanied by the discontinuous changes in the critical parameters~\cite{W.Gotze2009}. 

As the density increases and reaches the second minimum of $T_{g}(\rho)$ at $\rho \approx 1.8$, 
where the large particles form trimers and the small particles form dimer, 
the logarithmic relaxation 
is observed for the self part of the large particles, $\Phi_{s, L}(k,t)$,
and for the collective and self parts of the small particles, 
$\Phi_{S}(k,t)$ and $\Phi_{s, S}(k,t)$,  as shown in 
Figs.~\ref{fig:F_Fs} (e), (j), and (k).
$\Phi_{L}(k,t)$ remains to be a two-step relaxation function. 
At the third minimum at $\rho \approx 2.5$, where the number of particles per cluster increases further, 
we have not found any logarithmic relaxation any more.
Although the higher-order singularities are most clearly seen at the phase boundary from the monomer
to dimer phases, it is unclear whether they are 
completely absent at higher densities or they are simply masked 
by other complicated relaxation mechanism such as intercluster hopping.

\section{Conclusion}
\label{sec:conclusion}

\begin{figure}
 \begin{center}
	\includegraphics[width=1.0\columnwidth]{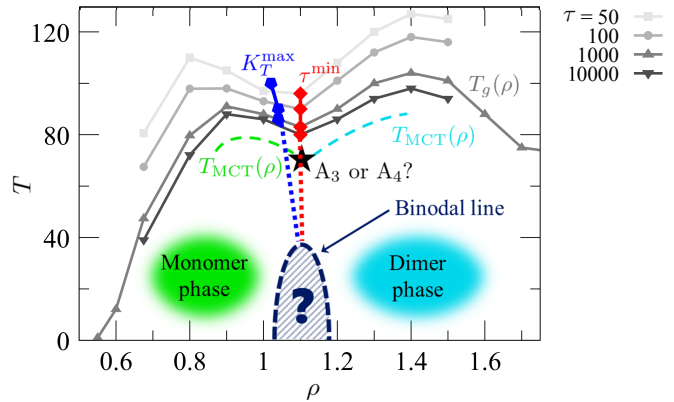}
 \end{center}
 \caption{
A schematic phase diagram near the glass-glass transition around $\rho\approx 1.1$.
 The black solid line is the glass-transition line, $T_{g}(\rho)$, defined as the iso-$\tau$ line with $\tau=10^3$.
	The gray solid lines are the iso-$\tau$ lines for different $\tau$ (see legend).
	The broken green and cyan lines are the speculated MCT critical temperatures $T_\text{MCT}$ 
	for the monomer and dimer glasses, respectively.
	The $\text{A}_3$ (or $\text{A}_4$) singular point (star) lies at the end point of one or both of the two lines.
	The blue pentagons and red diamonds are the loci of the maxima of the compressibility 
	and the minima of the iso-$\tau$ lines obtained by our simulations, respectively.
 	These two loci are extrapolated to lower temperatures 
 	where they may meet at the putative liquid-liquid critical point 
 	which is given by the terminal point of the coexistence region
 	drawn as the dark blue shaded area.
}
 \label{fig:schematic_phase_diagram}
\end{figure}

In this paper, we have presented detailed numerical results for structural and dynamical properties 
of the binary generalized Hertzian potential (GHP) fluids supercooled near the glass transition temperatures.
We especially focused on 
the interplay of glassy slow dynamics with cluster formation at high densities. 
The multiple maxima and minima observed in $T_{g}(\rho)$ synchronizes with the growth of the cluster
sizes.
The cluster size $n$ shows stepwise increases at every minimum of $T_g(\rho)$.
Contrary to the conventional glass formers, each of collective and self parts 
of the intermediate scattering functions for the large and small particles 
exhibits a distinct dynamics at high densities. 
The collective part is dominated by dynamics of the clusters' center of mass, 
whereas the self part is influenced sensitively by hopping and intra-cluster fluctuations.
The terminal relaxations for both the collective and self parts are influenced by the life time of the
clusters. 
Most significant in our studies is the logarithmic relaxation observed most clearly at the phase
boundaries separating the monomer and dimer glasses shown as the minima of $T_{g}(\rho)$. 
At the first minimum of $T_{g}(\rho)$, only the large particles exhibits the logarithmic relaxation, 
and the singularity of the small particles is observed at the second minimum.
The singular dynamics we observed agrees well with the prediction of the mode-coupling
theory~\cite{W.Gotze2002, W.Gotze2009}.  
However, the singularity were not clearly observed at higher densities where
the size of clusters are larger, which may be hidden due to the complex intra- and inter-cluster dynamics.
The stepwise change of $n$ as a function of density and the fact that this change becomes sharper as
the temperature is decreased suggest the putative thermodynamic liquid-liquid transition 
between the different cluster phases at an even lower temperature.
We also observed that the isothermal compressibility develops a peak in the vicinity of the density 
at which the first clustering is observed.
The temperature range over which we have investigated is too high to draw any solid conclusion about
the liquid-liquid phase transition. Further and independent investigation is required. 
However, we envisage that the (thermodynamics + dynamic) phase diagram near 
the first glass-glass transition (from the monomer to dimer glass phases) would look like 
Fig.~\ref{fig:schematic_phase_diagram}. 
As sketched in this figure, we speculate that the liquid-liquid phase binodal line lies down at the low temperature
region and would look similar to the solid-solid binodal found for the cluster crystal of the
monatomic Generalized Exponential Model (GEM) [see Fig. 1 of Ref.~\cite{K.Zhang2010}].
On the other hand, the MCT dynamical transition line would look like the two dashed
lines $T_\text{MCT}(\rho)$'s in this figure. 
We expect that at the intersection (more precisely at the close
vicinity of the intersection) of the two lines rests the A$_3$ or A$_4$ singular point. Note that
this is not a proof, but it is known in many systems that the higher order singularities are
observed in the vicinity of the intersection or termination point where the two glass lines
meet \cite{W.Gotze2009}. Also plotted in this figure are the loci of the maxima of the compressibility and
the minima of the iso-$\tau$ lines ($T_g(\rho)$). This is still a preliminary result but it suggests that
the two loci meet at a prospective liquid-liquid critical point at a low temperature.
We still do not know whether the observed higher order dynamical singularity is generically
connected to the underlying liquid-liquid thermodynamic transition. The glass-glass
dynamic crossover (such as the fragile-strong crossover) above the liquid-liquid critical point
has been discussed in the past but, to the best of our knowledge, the dynamical singularity
has never been discussed in the context of the liquid-liquid phase transition. It would be
worthwhile to study whether this dynamical anomaly is universally linked to thermodynamic
anomaly.

\begin{acknowledgments}
We thank D. Coslovich, A. Ikeda, H. Ikeda, and M. Ozawa for helpful discussions.
We acknowledge KAKENHI Grants 
No.~25103005, 
No.~25000002, 
and No.~16H04034. 
\end{acknowledgments}

%


\end{document}